\documentclass{ws-procs9x6}
\usepackage{array}

\begin{document}
\begin{center}
To appear on the Proceedings of the 13th ICATPP Conference on \\ 
Astroparticle, Particle, Space Physics and Detectors \\
for Physics Applications, \\
Villa Olmo (Como, Italy), 23--27 October, 2013, \\
to be published by World Scientific (Singapore)
\end{center}
\vspace{-2.0 cm}
\title{SCINTILLA \\ 
A European project for the development of scintillation detectors and new technologies for nuclear security}

\author{A. Alemberti$^2$, M. Battaglieri$^1$, E. Botta$^2$, R. DeVita$^1$, E. Fanchini$^{1,*}$ and G.Firpo$^2$ \\ 
On behalf of the SCINTILLA collaboration}

\address{
$^1$Istituto Nazionale di Fisica Nucleare (INFN), Sezione di Genova, Italy \\
$^2$Ansaldo Nucleare (ANN), Italy\\
$^*$E-mail: fanchini@ge.infn.it}

\begin{abstract}
Europe monitors transits using radiation detectors to prevent illicit trafficking of nuclear materials. The SCINTILLA project aims to develop a toolbox of innovative technologies designed to address different usage cases. This article will review the scope, approach, results of the first benchmark campaign and future plans of the SCINTILLA project. 
\end{abstract}


\bodymatter
\noindent 
\section{The homeland security surveillance}\label{sec_0}

The advantages of worldwide interconnectivity, with free movement of people and goods, is countered by the need to constantly monitor these movements in order to identify and prevent illegal trafficking of sensitive nuclear materials. There exists a broad variety of sensitive targets to survey, not only at international borders but also in hospitals, where radioactive isotopes are used; vehicle checkpoints within a city or nation; and places where large numbers of people gather, such as airports and public events. Both surveillance and prompt intervention systems are needed to ensure civil security is maintained.\\
The detection of Strategic Nuclear Material (SNM) relies on the type of emitted radiation. In most the cases the tools used are Radiation Portal Monitors (RPM) made of a combination of 2 detectors, one for neutron detection and one used for gamma detection and identification. Considering the case of plutonium and other fissile elements, the characteristic signature derives from the neutron emission, opposite to a large variety of common materials, which are gamma emitters. Examples are the substances known as Naturally Occurring Radioactive Materials (NORM), which are common in such goods as fruits or construction materials (bananas, coffee, concrete, fertilizers, etc.) and they have to be discriminated from other materials like medicals ($^{67}$Ga, $^{131}$I, etc.) or other gamma sources ($^{241}$Am, $^{60}$Co, etc.). These radiation sources cover a broad energy range from a few tens of eV up to few MeV, making gamma and neutron detection very challenging. In the case of neutron detection, devices used for nuclear security are mostly based on $^3$He technology. In the last few years the market trend of $^3$He encountered huge problems in matching supply and demand~[\cite{Kouzes09}]  causing an exponential increase in the price. For this reason, it become important to develop an alternative detection systems based on new technologies~[\cite{Kouzes10}].

\section{The SCINTILLA project}\label{sec_1}
It is almost impossible to cover the many variety of possible situations, described in Sec.~\ref{sec_1}, with a single detector. \\
The SCINTILLA project~[\cite{SCI}], within the European Community's Seventh Framework Program (FP7), is proposing an innovative and alternative way of addressing radiation detection in the nuclear security field, providing a toolbox of devices developed for specific usage cases. SCINTILLA involves a consortium of 9 partners, including 5 research groups (CEA, EK, Fraunhofer INT institute, INFN and Joint Research Center (JRC)) and 4 companies (ANN, Arttic, Saphymo and Symetrica), providing 9 detector prototypes listed in Tab.~\ref{table:1}. These devices are based on 6 different technologies (4 scintillator--based and 2 semiconductor  technologies) and cover 6 different usage cases that were defined within the scope of the project.
\begin{table}
\tbl{Parameters of the technologies under development.}
{\scriptsize
{\begin{tabular}{|>{\centering}m{3.4cm}|>{\centering}m{1.81cm}|>{\centering}m{2.3cm}|c|}
\hline \textbf{Detector}	& \textbf{Sensitivity}  & \textbf{Usage case} & \textbf{Developer} \\
\hline Gd--lined plastic scintillator   	& $n+\gamma$  	 			  & {\scriptsize container, vehicle} 				& INFN/ANN \\
\hline LiZnS  detector    					& $n$								  & {\scriptsize container, vehicle, luggage, people}		& Symetrica \\
\hline NaI(Tl) spectrometer     			& $\gamma$ {\scriptsize spectroscopy} & {\scriptsize luggage, people}			& Symetrica \\
\hline 2 PVT spectrometers 				& $\gamma$ {\scriptsize spectroscopy} & {\scriptsize container, vehicle, luggage, people	}		& Symetrica \\
\hline 2 PSD Plastic scintillators\footnotemark[1] & $n+\gamma$ &{\scriptsize container, vehicle, luggage, people	}       & CEA  and Saphymo \\
\hline CZT gamma camera 				& $\gamma$ 					  & {\scriptsize  portable device}				& CEA \\
\hline Mini--CZT  								& $\gamma$ 					  & {\scriptsize  miniature device}				& CEA \\
\hline
\end{tabular}}\label{table:1}
}
\end{table}
\footnotetext[1]{PSD: pulse shape discrimination}
For this reason, the SCINTILLA project is also providing a set of test--bed services and technology benchmarks to perform testing of detectors according to international standards and determine performance and improvements in technology along the full duration of the project.

\section{The first technology benchmark}\label{sec_2}
One of the most important aspect of the entire program is the evaluation of the state of development of the technologies, with the possibility to carry out dedicated radiological performance tests of single detectors or multiple devices in parallel.
The program foresees annual benchmarks for all partner's prototypes in one of the 3 partner facilities: JRC, in Italy, is specialized in radiometric dynamic tests of multiple devices according to international standard requirements; the hungarian EK facility is used  for measurements with radioactive sources and the Fraunhofer INT facility, in Germany, is in charge of final assessments of integrated systems.\\
The first benchmark was held in February 2013 at the JRC ITRAP+10 facility where all 7 RPM technologies were tested. The remaining 2 detectors, the CZT gamma camera and the mini--CZT device, will be tested during the second benchmark in 2014. The laboratory is equipped with a movable cart, carrying the radioactive source, which passes in front of the detectors under test, at a  programmable speed.  The testing conditions and in particular the source speed, detector--to--source distance, and source type were selected according to the usage cases the detectors are developed for, following international standards (IEC and ANSI)
~[\cite{ANSI_1, ANSI_2,IEC_1,IEC_2}]

\subsection{Benchmark results}
The first SCINTILLA benchmark was  an important milestone for the project. \\
Tested technologies can be divided into two groups: neutron detectors and gamma spectroscopy detectors. The 4 neutron devices, 3 based on plastic scintillators and one based on Lithium technology, detected all neutron source transits. For the 3 scintillator based detectors, the intrinsic capability to detect photons gave the possibility to test them also with gamma sources. Also in this case, their behaviour was very promising, especially for the detection of gamma of medium and high energies.\\ 
The second group of 3 gamma spectroscopy technologies were exposed to the same radioisotopes:  one detector was tested for gamma identification while the other 2, small plastic scintillators, for gamma categorization (\textit{NORM} or \textit{NOT NORM}), both showing positive results.\\
One of the RPM tested devices is the neutron detector developed in collaboration by INFN and ANN. It is based on a combination of plastic scintillator and Gadolinium, a material with high cross section for neutron capture ($\sigma_{nc} = 259 kb$ for thermal neutrons).   
The good combination and compromises between innovative design, robust and commercial components, leads to a alternative and versatile detector. Including the  intrinsic capability of gamma detection refined with a custom made event selection algorithm, transforms a neutron detector into a device with gamma discrimination in terms of energy ranges.
\begin{figure}[h!]
\begin{center}
\includegraphics[scale=0.48]{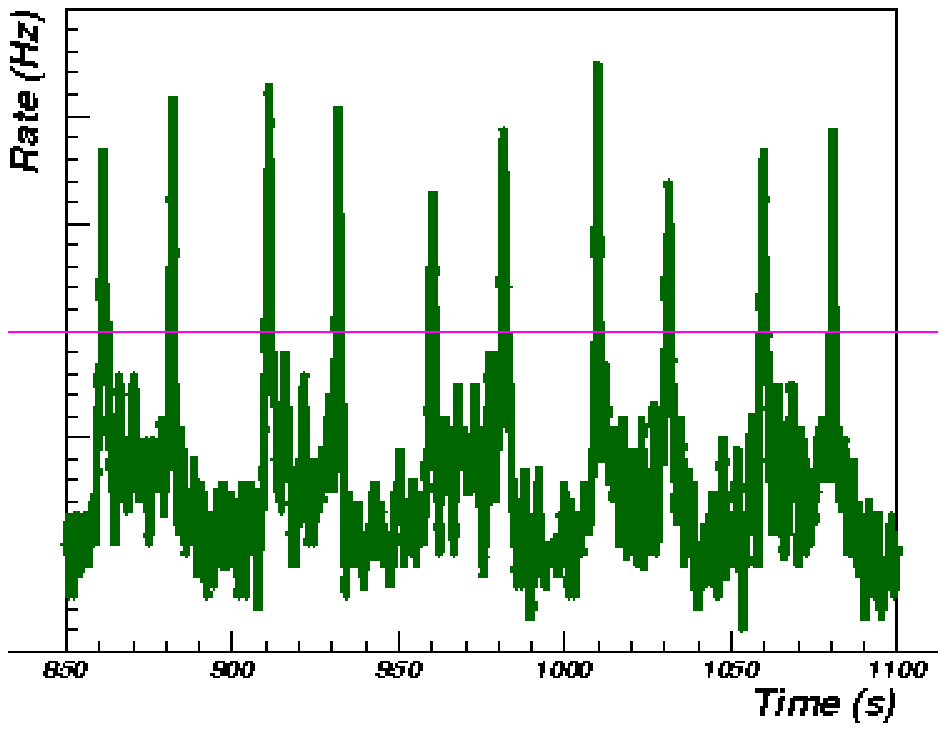}
\hspace{0.5cm}
\includegraphics[scale=0.48]{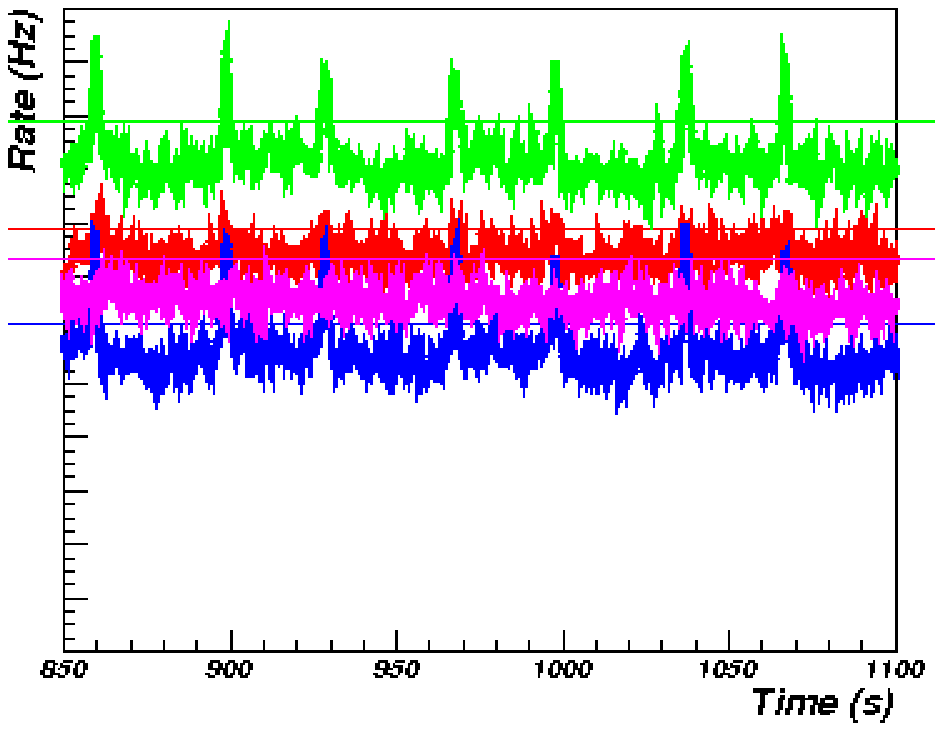}
 \caption{Zoom of two dynamic measurements taken from the INFN/ANN detector. The plot on the left shows 10 neutron source transits ($^{252}$Cf)  in front of the detector which correspond to the peaks. On the left, 7 passages of the $^{137}$Cs source are shown. Each coloured line represent a different energy range for energy discrimination.}\label{fig:1}
\end{center}
\end{figure}
Plots in Fig.~\ref{fig:1} show the measured rates as a function of time, for two dynamic measurements performed during benchmark. The straight lines  represent the alarm thresholds set and the peaks, which correspond to the increase of the detection rate, are due to source passages in front of the detector. The right plot of Fig.~\ref{fig:1} shows a $^{137}Cs$ measurement  and the left plot of a neutron measurement obtained with a $^{252}$Cf source. In the right plot, each coloured line corresponds to the gamma rate measured in a specific energy range, but only three of the 4 lines show peaks revealing all source transits. The magenta line,  corresponding to the highest  energy range well above the 661.7 keV photon energy of the $^{137}Cs$ decay, is flat and gives an indication of the maximum energy emission of the source under test.  All performed tests revealed very positive results but  also gave indications  for further improvements that will enhance the detection performances. These modifications are being implemented in  new versions of detector prototypes that will be ready for the second benchmark campaign of the early 2014. In summary, very positive and encouraging results were obtained confirming the high quality of the underlining technologies and of the instrument implementation.

\section{The RPM integration and the SCINTILLA network}\label{sec_3}
One of the objectives of  the  SCINTILLA program  is providing full RPM systems for different usage cases. Currently deployed RPMs are generally equipped with two types of radiation sensors, gamma and neutron detectors, combined with other devices, as annunciators and occupy sensors, managed by an intelligent system. This procedure is called integration. Even though all SCINTILLA detectors are developed independently, a common communication protocol was implemented and tested. This first integration test was held this summer, while the full systems  will be tested during the next benchmark campaign.  Next benchmarks are also open to partners of the Scintilla Partnership Network (\url{www.scintilla-project.eu}) which represents not only SCINTILLA partners but also outsiders (users, experts and companies) involved in the nuclear security field. The idea is to have a project with an impact in real life, not only to communicate the SCINTILLA project progresses, but also proposing new or different evaluation procedures to the ground for developments of novel technologies beyond the present state of art.

\bibliographystyle{ws-procs9x6}
\bibliography{ws-procs9x6}

\end{document}